\documentclass[12pt]{article}
\usepackage{graphicx}
\usepackage{amsmath}
\usepackage{amsfonts}
\usepackage{amssymb}
\newcommand{\be}{\begin{equation}}
\newcommand{\ee}{\end{equation}}
\newcommand{\bea}{\begin{eqnarray}}
\newcommand{\eea}{\end{eqnarray}}
\newcommand{\beq}{\begin{equation}}
\newcommand{\eeq}{\end{equation}}
\newcommand{\ba}{\begin{array}}
\newcommand{\ea}{\end{array}}
\newcommand{\beqa}{\begin{eqnarray}}
\newcommand{\eeqa}{\end{eqnarray}}

\newcommand{\cB}{{\cal B}}
\newcommand{\cO}{{\cal O}}

\begin{document}

\begin{flushright}%
April 2004 \\
hep-ph/0404127 \\
\end{flushright}

\vskip   1.5 true cm

\begin{center}
{\Large \textbf{The rare decay $K_{L} \rightarrow\pi^{0} \mu^{+} \mu^{-}$
      within the SM}} \\ [25 pt]
\textsc{Gino Isidori},${}^{1,2}$ \textsc{Christopher Smith},${}^{1}$ 
        and \textsc{Ren\'e Unterdorfer}${}^{1,3}$ \\ [20 pt]
${}^{1}~$\textsl{INFN, Laboratori Nazionali di Frascati, I-00044 Frascati,
      Italy} \\ [5 pt]
${}^{2}~$\textsl{Institut f\"ur Theoretische Physik, Universit\"at Bern, 
     CH-3012 Bern, Switzerland } \\ [5 pt]
${}^{3}~$\textsl{Institut f\"ur Theoretische Physik, Universit\"at
      Wien, A--1090 Wien, Austria } \\[25pt]

\textbf{Abstract}
\end{center}
\noindent
We present an updated analysis of the rare decay $K_{L}\rightarrow\pi^{0}%
\mu^{+}\mu^{-}$ within the Standard Model. In particular, we present the first
complete calculation of the two-photon CP-conserving amplitude within Chiral
Perturbation Theory, at the lowest non-trivial order. Our results confirm
previous findings that the CP-conserving contribution to the decay rate cannot
be neglected. By means of an explicit two-loop calculation, we show that this
contribution can be estimated with sufficient accuracy compared to the
CP-violating terms. We predict $\mathcal{B}(K_{L}\rightarrow\pi^{0}\mu^{+}%
\mu^{-})_{\mathrm{SM}}=(1.5\pm0.3)\times10^{-11}$, with approximately equal
contributions from the CP-conserving component, the indirect-CP-violating
term, and the interesting direct-CP-violating amplitude. The error of this
prediction is mainly of parametric nature and could be substantially reduced
with better data on the $K_{S}\rightarrow\pi^{0}\ell^{+}\ell^{-}$ modes. The
Standard Model predictions for various differential distributions and the
sensitivity to possible new-physics effects are also briefly discussed.

\setcounter{footnote}{0}
\vskip        1.0 true cm

\section{Introduction}

The rare decays $K_{L} \rightarrow\pi^{0} \ell^{+} \ell^{-}$ are among the
most interesting channels for precision studies of CP violation and flavour
mixing in the $\Delta S=1$ sector \cite{GW}. The decay amplitudes of these
processes have three main ingredients: \emph{i.}~a clean direct-CP-violating
component determined by short-distance dynamics; \emph{ii.}~an indirect-CP-violating 
(CPV) term due to $K_{L}$-$K_{S}$ mixing; \emph{iii.}~a
long-distance CP-conserving (CPC) component due to two-photon intermediate
states. Although generated by very different dynamics, these three components
are of comparable size within the Standard Model (SM). The precise knowledge
about their magnitude (particularly of \emph{ii.}~and \emph{iii.}) has
substantially improved in the recent past \cite{BDI}. This improvement has
been made possible by the observation of the $K_{S} \rightarrow\pi^{0}
\ell^{+} \ell^{-}$ decays \cite{NA48ee,NA48mm} and also by precise
experimental studies of the $K_{L} \rightarrow\pi^{0}\gamma\gamma$ diphoton
spectrum \cite{NA48_KLpgg,KTeV_KLpgg}.

The main difference of the $K_{L}\rightarrow\pi^{0}\mu^{+}\mu^{-}$ mode with
respect to the electron channel, which has recently been re-analysed in
Ref.~\cite{BDI}, is the two-photon CPC contribution. The latter corresponds to
transitions into final states where the lepton pair has $J^{\mathrm{CP}%
}=0^{++}$ or~$2^{++}$. The $J^{\mathrm{CP}}=0^{++}$ case can safely be
neglected in the electron mode, because of the strong helicity suppression
\cite{Se88,EPR88,FR89}. On the contrary, the $K_{L}\rightarrow\pi^{0}%
(\gamma\gamma)_{J=0}\rightarrow\pi^{0}(\ell^{+}\ell^{-})_{J=0}$ amplitude
generates the by far dominant CPC contribution in the muon mode
\cite{EPR88,Se93}. The main purpose of the present paper is a detailed
analysis of this CPC amplitude, which is a fundamental ingredient to obtain a
reliable SM prediction for the entire $K_{L}\rightarrow\pi^{0}\mu^{+}\mu^{-}$
decay width.

Chiral Perturbation Theory (CHPT) provides a natural framework to analyse the
CP-conserving $K_{L}\rightarrow\pi^{0}\mu^{+}\mu^{-}$ amplitude. Here the
first non-trivial contribution arises at $\mathcal{O}(p^{4})$ from two-loop
diagrams of the type $K_{L}\rightarrow\pi^{0}(P^{+}P^{-})\rightarrow\pi
^{0}(\gamma\gamma)\rightarrow\pi^{0}\mu^{+}\mu^{-}$, where $P=\pi,K$. This
structure is very similar to the two-loop $K_{S}\rightarrow\mu^{+}\mu^{-}$
amplitude analysed by Ecker and Pich~\cite{EP}: there are no counterterms at
$\mathcal{O}(p^{4})$ and the sum of all loop diagrams yields a finite
unambiguous result. The $K_{S}\rightarrow\mu^{+}\mu^{-}$ amplitude of
Ref.~\cite{EP} has indeed been used by Heiliger and Sehgal 
\cite{Se93} to estimate the dominant pion-loop terms 
in $K_{L}\rightarrow\pi^{0}\mu^{+}\mu^{-}$. These authors extracted 
from the $K_{S}\rightarrow\mu^{+}\mu^{-}$ amplitude an appropriate electromagnetic 
form factor,  describing the $(\pi^{+}\pi^{-})_{J=0} \rightarrow (\gamma\gamma)
\rightarrow \mu^{+}\mu^{-}$ transition, and used it in conjunction 
with a constant  $K_{L}\rightarrow 3\pi$ vertex to estimate 
the $K_{L}\rightarrow\pi^{0}\mu^{+}\mu^{-}$ amplitude. 
Although quite reasonable from a phenomenological point of view, 
this result is not fully justified within CHPT. 

By means of an explicit two-loop calculation, we show that the 
factorization of the electromagnetic form factor holds exactly 
at $\mathcal{O}(p^{4})$, independently of the structure of the  
$K_{L}\rightarrow 3\pi$ amplitude, and independently of the 
nature of the charged pseudoscalar meson inside the loop. 
We are then able to  present the first complete 
$\mathcal{O}(p^{4})$ estimate of the CPC
$K_{L}\rightarrow\pi^{0}\mu^{+}\mu^{-}$ 
amplitude, with the correct chiral dynamics 
of the $K_{L}\rightarrow 3\pi$ vertex and 
the missing kaon-loop contribution. 
In order to check the stability of this result,
we also performed a detailed study of possible higher-order
contributions, taking into account the recent experimental information on the
$K_{L}\rightarrow\pi^{0}\gamma\gamma$ decay distribution. 
The main outcome of this analysis is a reliable theoretical estimate 
of the ratio 
\begin{equation}
R_{\gamma\gamma} = 
 \frac{ \Gamma(K_{L}\rightarrow\pi^{0}\ell^{+}\ell^{-})_{\rm CPC}
  }{ \Gamma( K_{L}\rightarrow\pi^{0} \gamma\gamma) }~.
\label{eq:Rgg} 
\end{equation}
Despite numerator and denominator receive, separately, large corrections 
beyond $\mathcal{O}(p^{4})$, $R_{\gamma\gamma}$ turns 
out to be a rather stable quantity, with a theoretical 
error conservatively estimated to be around $30\%$.

Combining the analysis of the CPC amplitude with an updated 
estimate of the CPV terms, we finally arrive to the prediction 
 $\mathcal{B}(K_{L}\rightarrow\pi^{0}\mu^{+}
 \mu^{-})_{\mathrm{SM}} \approx 1.5 \times 10^{-11}$
for the total branching ratio. 
This turns out to be composed by almost equal amounts 
of  CPC, indirect-CPV and direct-CPV 
(including the interference) terms. The present error
of this estimate is around  $30\%$, 
but it is largely dominated by the parametric uncertainty due to the 
$\mathcal{B}(K_{S}\rightarrow\pi^{0}\ell^{+}\ell^{-})$ 
measurements. The irreducible theoretical uncertainty is 
around $10\%$ and can be further decreased by suitable 
kinematical cuts. We thus conclude that the  
$K_{L}\rightarrow\pi^{0}\mu^{+}\mu^{-}$ channel is a very 
promising mode for future precision studies of CP violation 
in the $\Delta S=1$ sector, with complementary virtues and 
disadvantages with respect to the other golden modes,
namely $K_{L}\rightarrow\pi^{0}\nu\bar\nu$ and  
$K_{L}\rightarrow\pi^{0}e^+e^-$. 
As an illustrative example of the discovery potential of the 
two $K_{L}\rightarrow\pi^{0} \ell^+\ell^-$ channels, we 
conclude our analysis showing that -- even with the present 
parametric uncertainties -- these two modes would yield 
a clear non-standard signature for the new-physics scenario 
recently proposed in Ref.~\cite{BFRS}.

The paper is organized as follows: in Section 2 we discuss the kinematics 
and the general amplitude decomposition of $K_{L}\rightarrow\pi^{0} \ell^+\ell^-$ 
decays. In Section 3 we update the prediction of the CP-violating 
rate for the muon mode. Section 4 contains the main results of this
work, namely the analytical and numerical study of the 
$K_{L}\rightarrow \pi^{0}(\ell^{+}\ell^{-})_{J=0}$ CP-conserving 
amplitude. The phenomenological analysis of 
$K_{L}\rightarrow\pi^{0} \mu^+\mu^-$ decays, within and beyond the SM,
is presented in  Section 5. The results are summarized in the 
Conclusions.

\section{Amplitude decomposition and differential distributions}

The most general decomposition of the $K_{L}\rightarrow\pi^{0}\ell^{+}\ell
^{-}$ amplitude involves four independent form factors ($S$, $P$, $V$, and
$A$) \cite{Agrawal}:
\begin{align}
&  \mathcal{A}(K_{L}\rightarrow\pi^{0}\ell^{+}\ell^{-})=\frac{e^{2}G_{F}%
}{(4\pi)^{2}}\Big[  m_{\ell}\bar{u}(p)v(p^{\prime})S(z,\tilde{y})+m_{\ell
}\bar{u}(p)\gamma_{5}v(p^{\prime})P(z,\tilde{y})\nonumber\\
&  \quad\qquad+(p_{K}+p_{\pi})_{\mu}\bar{u}(p)\gamma^{\mu}v(p^{\prime
})V(z,\tilde{y})+(p_{K}+p_{\pi})_{\mu}\bar{u}(p)\gamma^{\mu}\gamma
_{5}v(p^{\prime})A(z,\tilde{y})\Big]  ~.\qquad\label{eq:ff}%
\end{align}
The two independent kinematical variables can be chosen as
\begin{equation}
z=\frac{(p+p^{\prime})^{2}}{m_{K}^{2}}~,\qquad\tilde{y}=\frac{2}{m_{K}^{2}%
}\frac{p_{K}\!\cdot\!(p-p^{\prime})}{\beta_{\pi}(z)\beta_{\ell}(z)}%
~,\qquad\label{eq:yz}%
\end{equation}
where
\begin{equation}
r_{i}=\frac{m_{i}}{m_{K}}~,\quad\beta_{\ell}(z)=\left(  1-\frac{4r_{\ell}^{2}%
}{z}\right)  ^{1/2},\quad\beta_{\pi}(z)=\left(  1+r_{\pi}^{4}+z^{2}%
-2z-2r_{\pi}^{2}-2zr_{\pi}^{2}\right)  ^{1/2}.\
\end{equation}
The variable $\tilde{y}$ coincides with $\cos\theta_{\ell}$, where
$\theta_{\ell}$ is the angle between $K_{L}$ and $\ell^{-}$ spatial momenta in
the dilepton center-of-mass frame, and the two independent ranges are
\begin{equation}
4r_{\ell}^{2}\leq z\leq(1-r_{\pi})^{2}~,\qquad-1<{\tilde{y}}\equiv\cos
\theta_{\ell}<1~,
\end{equation}
With these conventions, the general expression of the differential decay rate
is
\begin{align}
&  \frac{d^{2}\Gamma}{dzd\tilde{y}}~=~\frac{\alpha^{2}G_{F}^{2}m_{K}^{5}%
\beta_{\pi}(z)\beta_{\ell}(z)}{2^{12}\pi^{5}}\Big\{  r_{\ell}^{2}\beta_{\ell
}^{2}(z)z|S(z,\tilde{y})|^{2}+r_{\ell}^{2}z|P(z,\tilde{y})|^{2}\nonumber\\
&  \qquad\quad+\beta_{\pi}^{2}(z)(1-\beta_{\ell}^{2}\tilde{y}^{2}%
)|V(z,\tilde{y})|^{2}+\left[  \beta_{\pi}^{2}(z)(1-\beta_{\ell}^{2}\tilde
{y}^{2})+4r_{\ell}^{2}(2+2r_{\pi}^{2}-z)\right]  |A(z,\tilde{y})|^{2}%
\nonumber\\
&  \qquad\quad+4r_{\ell}^{2}(1-r_{\pi}^{2})\mathrm{Re}\left[  A(z,\tilde
{y})^{\ast}P(z,\tilde{y})\right]  +4r_{\ell}^{2}\beta_{\pi}(z)\beta_{\ell
}(z)\tilde{y}\mathrm{Re}\left[  V(z,\tilde{y})^{\ast}S(z,\tilde{y})\right]
\Big\}  .\qquad
\end{align}

\medskip

The $\tilde y $ variable is CP-odd and leads to a strong kinematical
suppression. For this reason, it is convenient to expand the four form factors
in a multipole series:
\begin{equation}
F(z, \tilde y ) = F_{0}(z) + \tilde y F_{1}(z) + \ldots\qquad F=S,P,V,A~.
\end{equation}
The CP properties of the amplitudes are the following: the odd multipoles of
$S$ and the even multipoles of $P$, $V$ and $A$, vanish in the limit of exact
CP invariance. All short-distance contributions to the higher multipoles are
negligible, since they cannot be generated by dimension-six operators. To a
good approximation, we can also neglect all the long-distance contributions
generated by two-photon amplitudes in the CPV multipoles. As a result, we
are left with five potentially interesting terms:

\begin{description}
\item[$A_{0},P_{0}$] : direct CPV amplitudes of short-distance origin;

\item[$V_{0}$] : CPV amplitude which receives both short-distance (direct CPV
component) and long-distance (indirect CPV component) contributions \cite{EPR1,DEIP};

\item[$S_{0},V_{1}$] : CPC amplitude induced by the long-distance transition
$K_{L}\rightarrow\pi^{0}(\gamma\gamma)_{J=0,2}\rightarrow\pi^{0}\ell^{+}%
\ell^{-}$; with the $J^{\mathrm{PC}}=0^{++}$ two-photon state 
contributing only to $S_{0}$.\footnote{~Note that this decomposition 
of the amplitude in terms of Dirac structures and $\tilde{y}$ multipoles, 
which is quite useful to identify the dynamical origin of the various terms 
and their relative weight in the decay distribution, does not correspond 
to a projection on $|\ell^{+}\ell^{-}\rangle$ states of definite angular momentum. 
Thus the $S_{0}$ term receives both contributions from
$J^{\mathrm{PC}}=0^{++}$ and $J^{\mathrm{PC}}=2^{++}$ while, by construction, 
$V_{1}$ receives only contributions from $J^{\mathrm{PC}}=2^{++}$.}

\end{description}

As mentioned in the Introduction, these amplitudes turn out to be of
comparable magnitude despite they are generated by very different dynamical
mechanisms. Given the normalization in Eq.~(\ref{eq:ff}), the CPV
short-distance multipoles are of $\mathcal{O}(\mathrm{Im} (V_{ts}^{*}
V_{td})/e^{2} \sim10^{-3})$ within the SM, $V^{\mathrm{ind}}_{0}$ is of
$\mathcal{O}( \varepsilon\sim10^{-3})$ and the two-photon CPC terms are of
$\mathcal{O}(\alpha/(4\pi)\sim10^{-3})$. We stress that these five multipoles
are the only interesting terms also beyond the SM, and that only the
short-distance terms ($A_{0}$, $P_{0}$, $V^{\mathrm{dir}}_{0}$) could receive
sizable non-standard contributions.

\medskip

The differential $z$ distribution taking into account only the five
potentially interesting terms can be written as
\begin{equation}
\frac{d\Gamma}{dz}=\frac{d\Gamma_{\mathrm{CPC}}}{dz}+\frac{d\Gamma
_{\mathrm{CPV}}}{dz}~,
\end{equation}
where
\begin{align}
\frac{d\Gamma_{\mathrm{CPC}}}{dz}  &  =\frac{\alpha^{2}G_{F}^{2}m_{K}^{5}%
\beta_{\pi}(z)\beta_{\ell}(z)}{2^{11}\pi^{5}}\left\{  r_{\ell}^{2}\beta_{\ell
}^{2}(z)z|S_{0}(z)|^{2}+\frac{2}{15}\beta_{\pi}^{2}(z)\left(  1+\frac{6r_{\ell
}^{2}}{z}\right)  |V_{1}(z)|^{2}\right. \nonumber\\
&  \left.  \qquad\qquad+\frac{4}{3}r_{\ell}^{2}\beta_{\pi}(z)\beta_{\ell
}(z)\mathrm{Re}[S_{0}(z)V_{1}(z)^{\ast}]\right\}  ,\qquad\\
&  \qquad{}\nonumber\\
\frac{d\Gamma_{\mathrm{CPV}}}{dz}  &  =\frac{\alpha^{2}G_{F}^{2}m_{K}^{5}%
\beta_{\pi}(z)\beta_{\ell}(z)}{2^{11}\pi^{5}}\left\{  r_{\ell}^{2}%
z|P_{0}(z)|^{2}+\frac{2}{3}\beta_{\pi}^{2}(z)\left(  1+\frac{2r_{\ell}^{2}}%
{z}\right)  |V_{0}(z)|^{2}\right. \nonumber\\
&  \qquad\qquad+\left[  \frac{2}{3}\beta_{\pi}^{2}(z)\left(  1+\frac{2r_{\ell
}^{2}}{z}\right)  +4r_{\ell}^{2}(2+2r_{\pi}^{2}-z)\right]  |A_{0}%
(z)|^{2}\nonumber\\
&  \left.  \qquad\qquad+4r_{\ell}^{2}(1-r_{\pi}^{2})\mathrm{Re}\left[
A_{0}(z)^{\ast}P_{0}(z)\right]  \frac{{}}{{}}\right\}  ~. \label{eq:Gamma_CPV}%
\end{align}
The asymmetric integration on $\tilde{y}$ give rise to the following
CPV distribution
\bea
\frac{d A_{\mathrm{FB}}}{dz}  &=& \int d\tilde{y}~\frac{d^{2}\Gamma}{dzd\tilde{y}%
}~\mathrm{sgn(\tilde{y})}~=~\frac{\alpha^{2}G_{F}^{2}m_{K}^{5}\beta_{\pi}%
^{2}(z)\beta_{\ell}(z)}{2^{12}\pi^{5}}\nonumber\\
&&  \times\mathrm{Re}\left[  V_{0}(z)^{\ast}\left(  4r_{\ell}^{2}\beta_{\ell
}(z)S_{0}(z)+\beta_{\pi}(z)\left(  1+\frac{4r_{\ell}^{2}}{z}\right)
V_{1}(z)\right)  \right]~.
\eea
With a proper normalization, $A_{\mathrm{FB}}$ 
can be identified with the forward-backward 
or the energy asymmetry of the two leptons.

\section{The CP-violating rate}

\label{sect:CPV}

An updated discussion of the CP-violating amplitude for the electron case
($\ell=e$) can be found in Ref.~\cite{BDI}, to which we refer for more
details. Most of the ingredients of this analysis can easily be transferred to
the muon case. Following the notation of Ref.~\cite{BDI,DEIP}, the dominant
indirect-CPV amplitude leads to
\begin{equation}
V_{0}^{\mathrm{ind}}(z)=\pm\varepsilon\left[  a_{S}+b_{S}z+\frac{W_{S}^{\pi
\pi}(z)}{G_{F}m_{K}^{2}}\right]  \approx\pm\varepsilon(a_{S}+b_{S}z)~,
\end{equation}
where the $\pm$ refers to the relative sign between this long-distance term
and the short-distance (direct-CPV) component of the vector
amplitude.\footnote{~The separation of $V^{\mathrm{ind}}$ and $V^{\mathrm{dir}%
}$ is phase-convention dependent. We adopt the standard CKM phase convention
where, to a good approximation, we can set $\varepsilon=|\varepsilon
|e^{i\pi/4}$.} The latter is given by
\begin{equation}
V_{0}^{\mathrm{dir}}(z)=i\frac{4\pi y_{7V}}{\sqrt{2}\alpha}f_{+}%
(z)\mathrm{\operatorname{Im}}\lambda_{t}~,
\end{equation}
where, as usual, $\lambda_{t}$ denotes the CKM combination $V_{ts}^{\ast
}V_{td}$, $f_{\pm}(z)$ are the $K\rightarrow\pi$ form factors of the $\bar
{s}\gamma^{\mu}d$ current, and $y_{7V,A}$ are the Wilson coefficients of the
leading flavour-changing neutral-current (FCNC) operators \cite{BLMM}. Similarly, the axial-current 
operator leads to the clean short-distance terms
\begin{align}
A_{0}(z)  &  =i\frac{4\pi y_{7A}}{\sqrt{2}\alpha}f_{+}%
(z)\mathrm{\operatorname{Im}}\lambda_{t}~,\\
P_{0}(z)  &  =-i\frac{8\pi y_{7A}}{\sqrt{2}\alpha}f_{-}%
(z)\mathrm{\operatorname{Im}}\lambda_{t}~.
\end{align}
Note that, contrary to the electron case, here the pseudoscalar amplitude is
not helicity suppressed (see Eq.~(\ref{eq:Gamma_CPV})).

Thanks to the recent NA48 results,
\begin{align}
\cB(K_{S}\rightarrow\pi^{0}e^{+}e^{-})_{m_{ee}>165~\mathrm{MeV}}  &  =\left(
3.0_{-1.2}^{+1.5}\pm0.2\right)  \times10^{-9}~\cite{NA48ee}~,\\
\cB(K_{S}\rightarrow\pi^{0}\mu^{+}\mu^{-})  &  =\left(  2.9_{-1.2}^{+1.4}%
\pm0.2\right)  \times10^{-9}~\cite{NA48mm}~,
\end{align}
the large theoretical uncertainty about the size of the indirect-CPV amplitude
has been strongly reduced. Under the natural and mild assumption that the form
factor of the $K_{S}\rightarrow\pi^{0}\ell^{+}\ell^{-}$ amplitude is equal to
$f_{+}(z)$, i.e.~assuming $(b_{S}/a_{S})z=f_{+}(z)-1=0.39z$~\cite{DEIP}, 
the combination of these two measurements leads to
\begin{equation}
|a_{S}|=1.2\pm0.2~. \label{eq:as}
\end{equation}
Within the SM, the main parametric uncertainty on the direct-CPV amplitude is
due to $\mathrm{Im}\lambda_{t}$. According to recent global CKM fits
\cite{CKM}:
\begin{equation}
\mathrm{\operatorname{Im}}\lambda_{t}=\mathrm{\operatorname{Im}}(V_{ts}^{\ast
}V_{td})\overset{_{\mathrm{SM}}}{\longrightarrow}(1.36\pm0.12)\times
10^{-4}~.
\end{equation}
Keeping the explicit dependence on these two main input values, and combining
all the CPV contributions, we obtain the following phenomenological
expression:
\begin{equation}
\cB(K_{L}\rightarrow\pi^{0}\mu^{+}\mu^{-})_{\mathrm{CPV}}\,=\,10^{-12}%
\times\left[  3.7|a_{S}|^{2}\pm1.6|a_{S}|\left(  \frac{\displaystyle
\mathrm{\operatorname{Im}}\lambda_{t}}{\displaystyle  10^{-4}}\right)
+1.0\left(  \frac{\displaystyle  \mathrm{\operatorname{Im}}\lambda_{t}%
}{\displaystyle  10^{-4}}\right)  ^{2}\right]  ~. \label{eq:cpvt}%
\end{equation}
The numerical coefficients have been obtained setting $y_{7A}=-(0.68\pm
0.03)\alpha(M_{Z})$ and $y_{7V}=(0.73\pm0.04)\alpha(M_{Z})$, corresponding to
${\overline{m}}_{t}(m_{t})=167\pm5$~GeV and a renormalization scale between
$0.8$ and $1.2$~GeV~\cite{BLMM}. Using these inputs, the error on each of the
three coefficients in Eq.~(\ref{eq:cpvt}) is below $10\%$. The result depends
very mildly on $f_{-}$, which has been set to $f_{-}(z)=-0.14$; however, the
relative weight of the three terms in (\ref{eq:cpvt}) is different with
respect to the electron case. This happens because of the $r_{\ell}$ terms in
Eq.~(\ref{eq:Gamma_CPV}), which enhance the axial-current direct-CPV
contribution. As discussed in Ref.~\cite{BDI}, the preferred theoretical
choice for the sign of the interference term is the \emph{positive} sign.
The same conclusion has recently been reached also by Friot, 
Greynat and de Rafael, who reinforced this theoretical argument 
with a detailed dynamical analysis of $K\to \pi\gamma \to\pi \ell^+\ell^- $  
amplitudes in the large $N_C$ limit \cite{GD}. 

\section{The CP-conserving rate}

\label{sect:CPC} As discussed in the Introduction, the leading CHPT
contribution to the CP-conserving $K_{L}\rightarrow\pi^{0}\ell^{+}\ell^{-}$
amplitude is generated by two-loop diagrams of $\mathcal{O}(e^{4}p^{4})$. The
general structure of these diagrams is indicated in Fig.~\ref{fig:twoloops1}%
--\ref{fig:twoloops2}. Due to the absence of local contributions, the complete
two-loop amplitude of $\mathcal{O}(e^{4}p^{4})$ is finite and does not depend
on unknown coefficients.

\subsection{General decomposition of the loop amplitude}

\begin{figure}[t]
\begin{center}%
\[
\includegraphics[width=12cm]{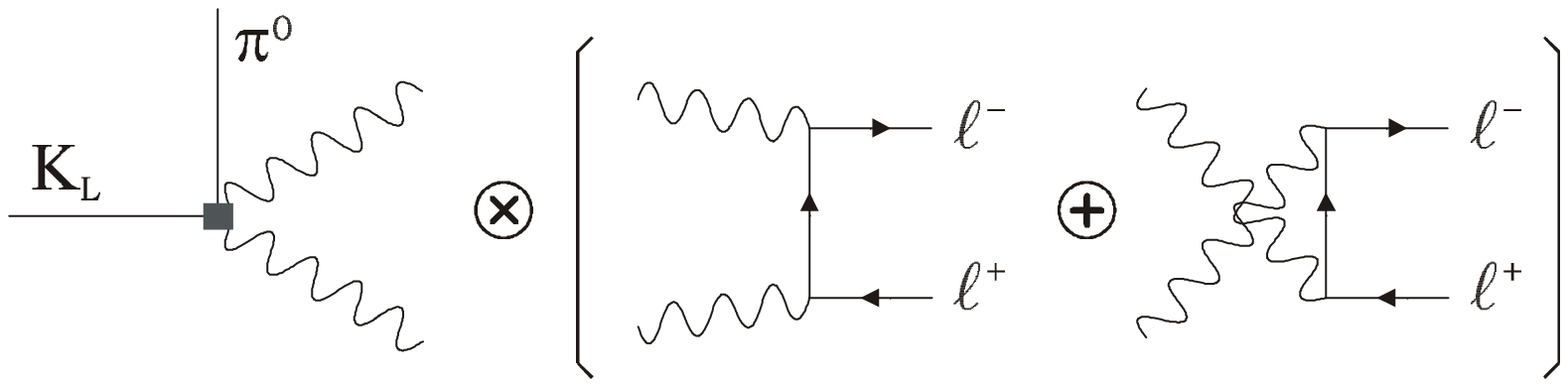}  \hspace{1cm}%
\]
\end{center}
\caption{General decomposition of the $K_{L}\rightarrow\pi^{0}\ell^{+}\ell
^{-}$ amplitude.}%
\label{fig:twoloops1}%
\begin{center}%
\[
\includegraphics[width=12cm]{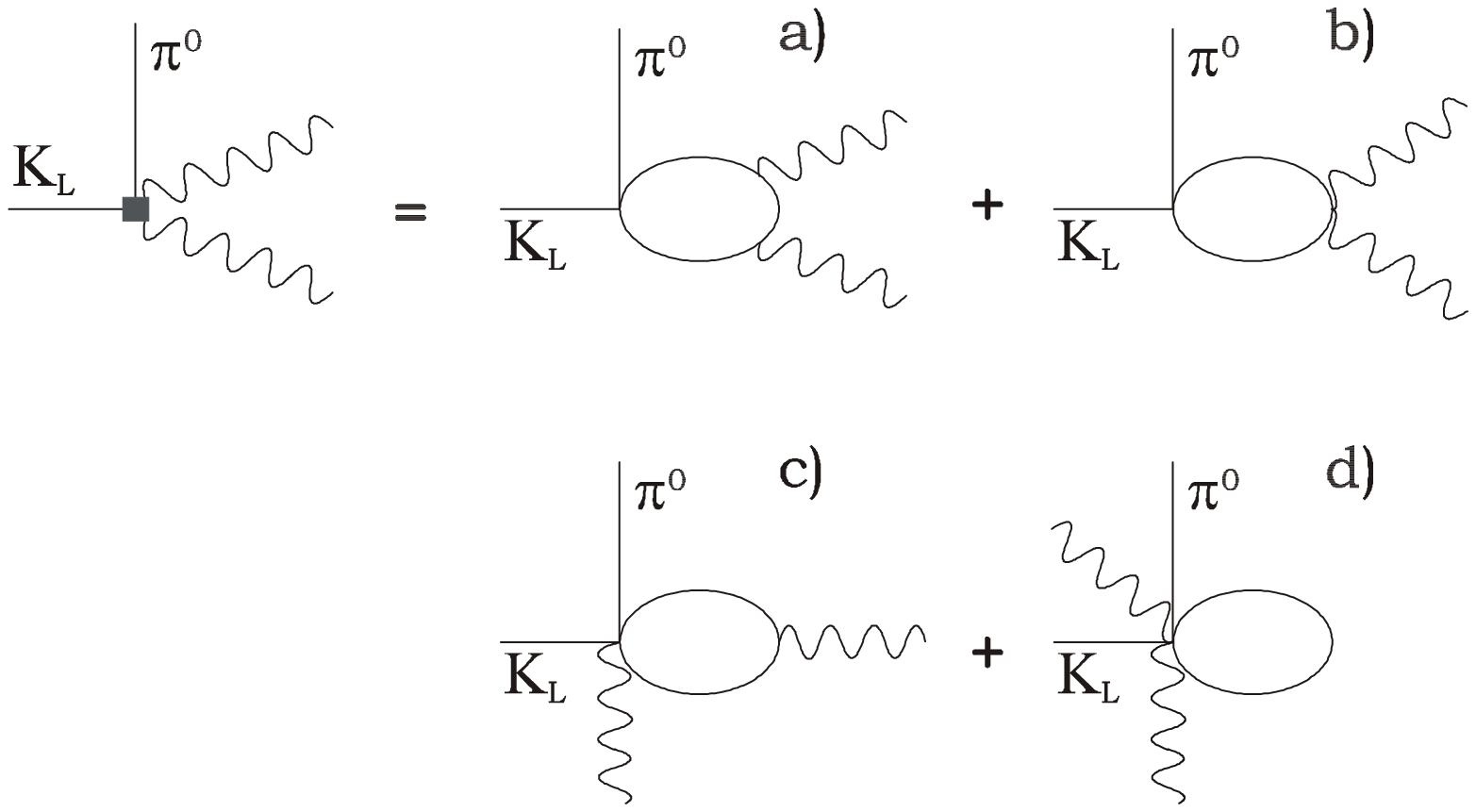}  \hspace{1cm}%
\]
\end{center}
\caption{Lowest-order diagrams contributing to the $K_{L}\rightarrow\pi
^{0}\gamma\gamma$ amplitude in the basis of Ref.~\cite{EPRgg}.}%
\label{fig:twoloops2}%
\end{figure}

{}From the general structure in Fig.~\ref{fig:twoloops1}, we can decompose the
CPC amplitude as
\begin{equation}
\mathcal{A}(K_{L}\rightarrow\pi^{0}\ell^{+}\ell^{-})_{\mathrm{CPC}%
}=\frac{ie^{2}}{2}\int\frac{d^{d}q^{\prime}}{(2\pi)^{d}}~A_{\gamma\gamma}%
^{\mu\nu}(p^{\prime}-q^{\prime},p+q^{\prime})~L_{\mu\nu}(p,p^{\prime
};q^{\prime})~, \label{eq:full}%
\end{equation}
where
\begin{equation}
L^{\mu\nu}(p,p^{\prime};q^{\prime})=\bar{u}(p)\left[  \gamma^{\nu}%
\frac{1}{-q^{\prime}\!\!\!\!/-m_{\ell}}\gamma^{\mu}+\gamma^{\mu}%
\frac{1}{q^{\prime}\!\!\!\!/+p\!\!\!/-p^{\prime}\!\!\!\!/-m_{\ell}}\gamma
^{\nu}\right]  v(p^{\prime})
\end{equation}
and $A_{\gamma\gamma}^{\mu\nu}$ is defined by the $K_{L}\rightarrow\pi
^{0}\gamma\gamma$ amplitude with off-shell photons:
\begin{equation}
\mathcal{A}[K_{L}\rightarrow\pi^{0}\gamma(\epsilon_{1},t)\gamma(\epsilon
_{2},k)]=A_{\gamma\gamma}^{\mu\nu}(t,k)\epsilon_{1}^{\nu}\epsilon_{2}^{\mu}~.
\end{equation}
A strong simplification arises by the observation that the leptonic tensor
obeys the Ward-identity relations
\begin{equation}
t_{\nu}L^{\mu\nu}=k_{\mu}L^{\mu\nu}=0~.
\end{equation}
This implies that we can neglect in $A_{\gamma\gamma}^{\mu\nu}$ all the
manifestly off-shell terms of the type $t^{\mu}t^{\nu}$, $k^{\mu}t^{\nu}$,
$k^{\mu}k^{\nu}$. The non-trivial Lorentz structure of $A_{\gamma\gamma}%
^{\mu\nu}$ is then identical to the one of the on-shell amplitude
\cite{EPRgg,CD} and its calculation proceeds in a very similar way.

As shown in Ref.~\cite{EPRgg}, employing a suitable basis for the pseudoscalar
meson fields, it is possible to reduce the number of independent diagrams
contributing to the $K_{L}\rightarrow\pi^{0}\gamma\gamma$ amplitude to the
four terms in Fig.~\ref{fig:twoloops2}. We note here that this structure can
be further simplified with a suitable decomposition of the $K\rightarrow
\pi^{0}P^{+}P^{-}$ amplitude, which acts as an effective vertex in the
diagrams \ref{fig:twoloops2}a--\ref{fig:twoloops2}b. At $\mathcal{O}(p^{2})$
we can write:
\begin{equation}
\mathcal{A}[K\rightarrow\pi^{0}P^{+}(p_{+})P^{-}(p_{-})]=i\left[  a_{0}%
+a_{1}z+a_{2}(p_{+}^{2}+p_{-}^{2}-2m_{P}^{2})\right]  ~, \label{eq:A3p}%
\end{equation}
where $z$ is defined as in Eq.~(\ref{eq:yz}) and the $a_{i}$ do not depend on
the meson momenta. At this order gauge invariance forces the effective
vertices of the diagrams \ref{fig:twoloops2}c--\ref{fig:twoloops2}d to be
completely determined by Eq.~(\ref{eq:A3p}) via the minimal substitution. It
is then easy to realize that the complete sum of the four terms in
Fig.~\ref{fig:twoloops2} is equivalent to the sum of \ref{fig:twoloops2}%
a--\ref{fig:twoloops2}b performed without the off-shell term $a_{2}$. We are
therefore left with the calculation of only two irreducible diagrams, with an
effective $K\rightarrow\pi^{0}P^{+}P^{-}$ vertex which does not depend on the
loop momenta:
\begin{equation}
M_{\pi^{0}PP}^{(2)}(z)=a_{0}+a_{1}z~.
\end{equation}
This re-organization of the calculation leads to several advantages: i) a
reduction of the number of relevant diagrams; ii) a manifest link with the
$K_{S}\rightarrow\ell^{+}\ell^{-}$ calculation of Ref.~\cite{EP}; iii) the
possibility to include a well-defined class of higher-order contributions (see
e.g.~Ref.~\cite{DEIN}). As we shall show later on, the latter step is achieved
in the case of the dominant pion loops by the replacement of $M_{3\pi}%
^{(2)}(z)$ with the $K\rightarrow\pi^{0}\pi^{+}\pi^{-}$ amplitude determined
by experiments.

The explicit integration over the $P$-meson loop in $A^{\mu\nu}_{\gamma\gamma
}$, namely
\[
A^{\mu\nu}_{\gamma\gamma} ( p^{\prime}- q^{\prime}, p + q^{\prime})^{(P)} = -
e^{2} M^{(2)}_{\pi^{0}PP}(z) \int\frac{ d^{d} q }{ (2\pi)^{d} } \frac{
(2q-k)^{\mu}(2q+t)^{\nu}- g^{\mu\nu} (q^{2} -m_{P}^{2}) }{ [(q+t)^{2}
-m_{P}^{2}] [(q-k)^{2} -m_{P}^{2}] [q^{2} - m_{P}^{2}] }~,
\]
leads to
\begin{align}
&  \mathcal{A}( K_{L} \to\pi^{0} \ell^{+} \ell^{-})_{\mathrm{CPC}}^{(P)} = -
\frac{e^{4}}{ 16\pi^{2} } M^{(2)}_{\pi^{0}PP}(z)\nonumber\\
&  \qquad\times\int_{0}^{1} dx \int_{0}^{1-x} dy \int\frac{ d^{d} q^{\prime}}{
(2\pi)^{d} } \frac{ u( p ) \gamma^{\nu}( - q^{\prime}\!\!\!\!/ -m_{\ell})
\gamma^{\mu}v( p^{\prime}) }{ [( q^{\prime})^{2}-m^{2}_{\ell}]( q^{\prime}+ p
)^{2}( q^{\prime}- p^{\prime})^{2} }\nonumber\\
&  \qquad\times\left[  \frac{ 4 xy}{ \Delta(x,y)} t_{\mu}k_{\nu}+ \frac{ -2xy
(z m_{K}^{2}-t^{2}-k^{2}) +y(2y-1)t^{2} +x (2x-1)k^{2} }{ \Delta(x,y)}
g_{\mu\nu} \right]  ~, \qquad
\end{align}
where
\begin{equation}
\Delta(x,y) = m_{P}^{2} +i\epsilon-xy(z m_{K}^{2}) +t^{2} (-1+x+y) +k^{2}
(-1+x+y)~.
\end{equation}

\subsection{Complete analytical expression and numerical integration}

The second loop integration is definitely more involved, but can still be
expressed in a compact way in terms of four Feynman parameters. Using the
equation of motion on the external lepton fields we can write
\begin{equation}
\mathcal{A}(K_{L}\rightarrow\pi^{0}\ell^{+}\ell^{-})_{\mathrm{CPC}}%
^{(P)}=\frac{\alpha^{2}}{2\pi^{2}}M_{\pi^{0}PP}^{(2)}(z)\frac{m_{\ell
}{\bar u}(p)v(p^{\prime})}{zm_{K}^{2}}\mathcal{I}\left(  \frac{r_{l}^{2}}%
{z},\frac{r_{P}^{2}}{z}\right)  ~,
\label{eq:fullp2}
\end{equation}%
\begin{align}
\mathcal{I}\left(  a,b\right)   &  =\int_{0}^{1}dx\int_{0}^{1-x}ydy\int
_{0}^{1}dw\int_{0}^{1}du\frac{1}{\alpha+\beta u-i\varepsilon}\nonumber\\
&  \times\left\{  \left(  2x+2y-1\right)  \left[  1+u\left(  1-w\right)
\right]  -\frac{x}{1+u\gamma+i\varepsilon}\right\}  ~, \label{eq:Iab}%
\end{align}
where
\begin{align}
\alpha &  =w\left[  b+y\left(  y-1\right)  \right]  ~,\qquad\qquad
\gamma=-1+w\frac{x\left(  1-x\right)  -b}{\left(  1-x-y\right)  \left(
x+y\right)  }~,\nonumber\\
\beta &  =wy\left(  1-\frac{wx}{x+y}\right)  \left(  1-x-y\right)  +a\left(
1-w\right)  ^{2}\left(  x+y\right)  \left(  1-x-y\right)  ~. \label{eq:alpha}%
\end{align}
The notation of Eqs.~(\ref{eq:Iab}) and (\ref{eq:alpha}) has been chosen in
order to show the perfect analogy and the agreement with the results of
Ref.~\cite{EP}, in the appropriate kinematical limit. As discussed by Ecker
and Pich, the integral (\ref{eq:Iab}) has three possible absorptive cuts: the
$\gamma\gamma$ cut, and the $PP$ and $PP\gamma$ cuts (both open only for
$b<1/4$). The real contribution associated to the double $PP$--$\gamma\gamma$
cut turns out to be vanishing.

\begin{figure}[t]
\begin{center}%
\[
\includegraphics[width=12cm]{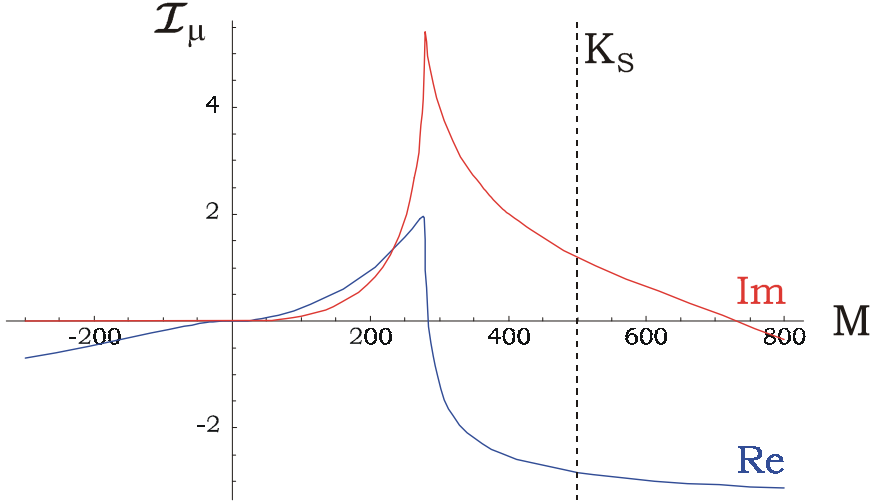}  \hspace{1cm}%
\]
\end{center}
\caption{Behavior of $\mathcal{I}(  m_{\mu}^{2}/M^{2},m_{\pi}^{2}/M^2)$ 
as a function of $M$ (in MeV).}
\label{fig:IabMu}%
\end{figure}

\begin{figure}[t]
\begin{center}%
\[
\includegraphics[width=12cm]{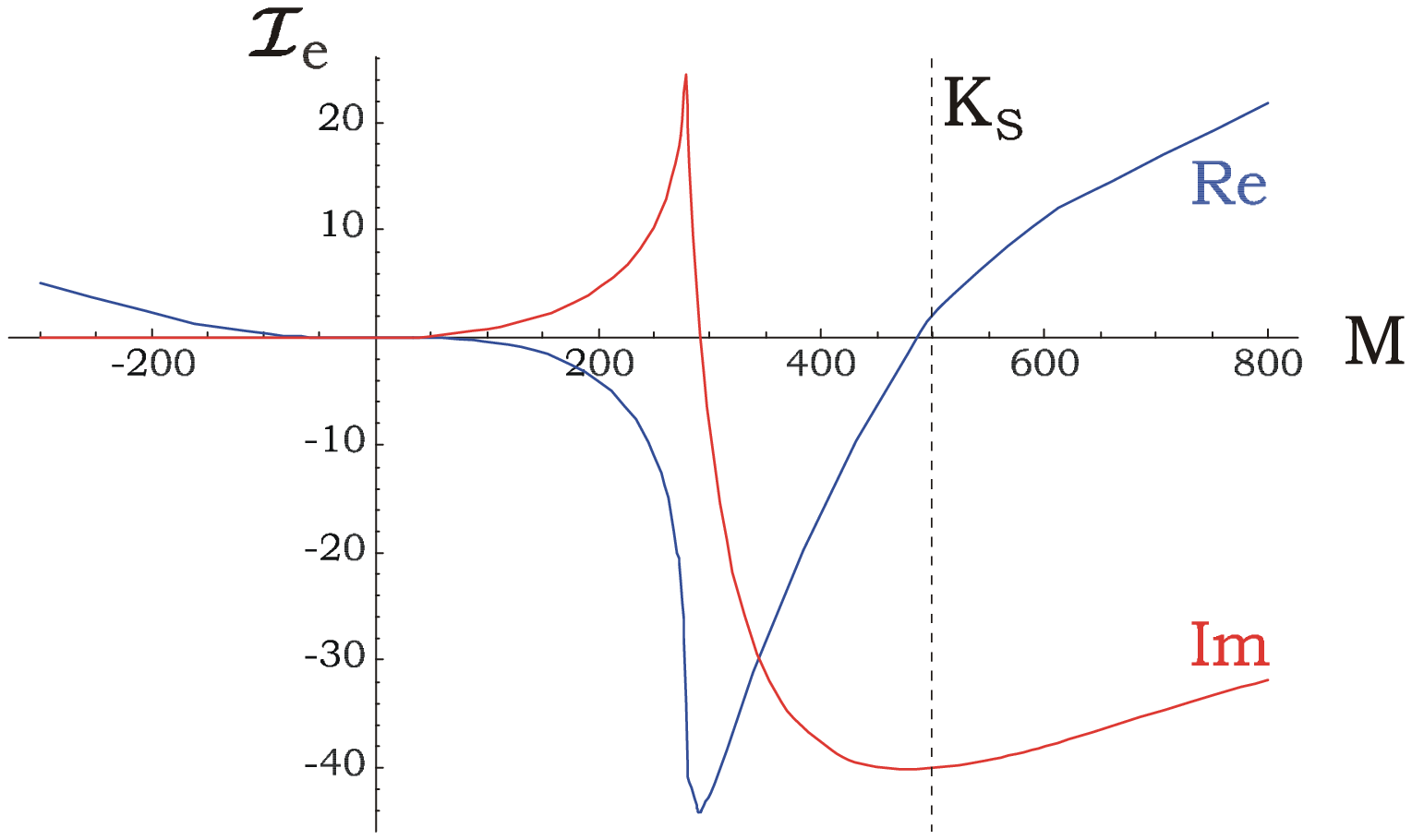}  \hspace{1cm}%
\]
\end{center}
\caption{Behavior of $\mathcal{I}(  m_{e}^{2}/M^{2},m_{\pi}^{2}/M^2)$ 
as a function of $M$ (in MeV).}%
\label{fig:IabEl}%
\end{figure}

These analytic properties become manifest performing the integration over the
$u$ variable in Eq.~(\ref{eq:Iab}). This leads to
\bea
\mathrm{Re}~\mathcal{I}(a,b)  &&  =\displaystyle  \int_{0}^{1}dx\int_{0}%
^{1-x}ydy\int_{0}^{1}dw\left\{  \frac{x}{\alpha\gamma-\beta}\log\left|
\frac{\alpha+\beta}{\alpha\left(  1+\gamma\right)  }\right|  \right.
\nonumber\\
&&  \qquad\qquad\left.  +\frac{2x+2y-1}{\beta}\left(  1-w+\left(
1-\frac{\alpha}{\beta}\left(  1-w\right)  \right)  \ln\left|  \frac{\alpha
+\beta}{\alpha}\right|  \right)  \right\}  ~,\label{ReIab}\\
\mathrm{Im}~\mathcal{I}(a,b)  &&  =\displaystyle  \pi\int_{0}^{1}dx\int
_{0}^{1-x}ydy\int_{0}^{1}dw\frac{2x+2y-1}{2\beta}\left(  1-\frac{\alpha}%
{\beta}\left(  1-w\right)  \right)  \left(  \frac{\left|  \alpha+\beta\right|
}{\alpha+\beta}-\frac{\left|  \alpha\right|  }{\alpha}\right) \nonumber\\
&&  \qquad\qquad\left.  +\lim_{\varepsilon\rightarrow0}\frac{x\left(
\alpha\gamma-\beta\right)  }{\left(  \alpha\gamma-\beta\right)  ^{2}%
+\varepsilon^{2}}\frac{1}{2}\left(  \frac{\left|  \alpha+\beta\right|
}{\alpha+\beta}-\frac{\left|  \alpha\right|  }{\alpha}+\frac{\left|
1+\gamma\right|  }{1+\gamma}-1\right)  \right\}  ~. \label{ImIab}%
\eea
The above representation, which holds independently of the possible signs of
$\alpha$, $\beta$ and $\gamma$, takes into account all the discontinuity
prescriptions and is suitable for numerical integration. This last
step is actually quite delicate because, though their combination is finite,
both the $PP$ and the $PP\gamma$ cuts are infrared  
divergent (the electron case is especially difficult 
since collinear singularities are much stronger). This means
that large cancellations occur between the various terms of $\mathcal{I}%
(a,b)$. We used a numerical integration routine based on VEGAS \cite{Vegas},
implemented in \textit{Mathematica} \cite{Math} on which we relied for
preventing loss of numerical significance. The results are shown in
Figs.~\ref{fig:IabMu}--\ref{fig:IabEl} for $P=\pi$ (where $M<0$ should be
understood as $i\left|M\right|$). The behavior for $P=K$ is entirely
determined by the $\gamma\gamma$ cut and it is rather smooth 
(as can be guessed from
Figs.~\ref{fig:IabMu}--\ref{fig:IabEl} below $2m_{\pi}$).

The values at $z=1$, relevant for $K_{S}\rightarrow\pi^{+}\pi^{-}%
\rightarrow\gamma\gamma\rightarrow\ell^{+}\ell^{-}$, are%
\begin{equation}
\begin{tabular}
[c]{ll}%
$\mathcal{I}\left(  r_{e}^{2},r_{\pi}^{2}\right)  =2.11-40.41i\;\;\;\;\;\;$ &
$\left(  \pm0.15\right)  ~,$\\
$\mathcal{I}\left(  r_{\mu}^{2},r_{\pi}^{2}\right)  =-2.821+1.216i$ & $\left(
\pm0.001\right)  ~,$%
\end{tabular}
\end{equation}
which are compatible with the figures quoted in Ref.~\cite{EP}. 
To further check our results, we have taken
advantage of the fact that $\mathcal{A}(J^0\left(  s\right)  \rightarrow
\pi^{0}\ell^{+}\ell^{-})_{\mathrm{CPC}}$ satisfies an unsubtracted dispersion
relation. In particular, the reduced amplitude 
$\mathcal{I}\left(M^{2}\right)  \equiv\mathcal{I}(m_{\ell}^{2}/M^{2},m_{P}^{2}/M^{2})$
obeys the relation 
\begin{equation}
\operatorname{Re}\frac{\mathcal{I}\left(  M^{2}\right)  }{M^{2}}=\frac{P}{\pi
}\int_{0}^{\infty}\frac{ds}{s-M^{2}}\operatorname{Im}\frac{\mathcal{I}\left(
s\right)  }{s}~. \label{DispRel}%
\end{equation}
After having computed the imaginary part over a large range of $s$ from
Eq.~(\ref{ImIab}), we obtained a very good agreement between the real part
values found by numerical integration of (\ref{DispRel}) or directly from (\ref{ReIab}),
for both $M^{2}>0$ and $M^{2}<0$. This check is particularly significant in
the region $M^{2}<0$, where the integration of (\ref{ReIab}) is stable, 
being singularity-free, while (\ref{DispRel}) remains very sensitive to the
cancellations among $PP$ and the $PP\gamma$ cuts in $\operatorname{Im}%
\mathcal{I}(s)$.

\subsection{Total and differential rate, and inclusion of higher-order terms}
\begin{figure}[t]
\begin{center}%
\[
\includegraphics[width=9.5cm]{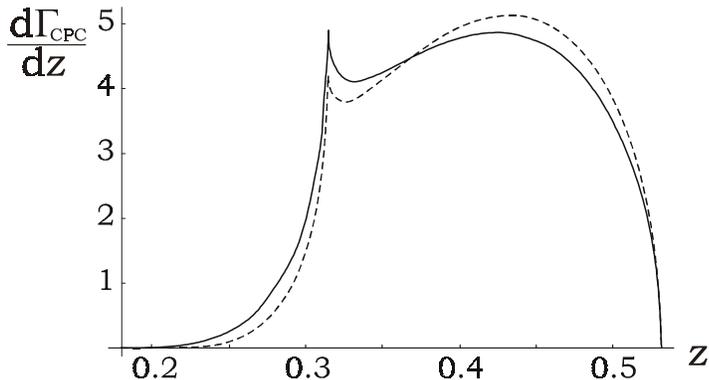}  \hspace{1cm}%
\]
\end{center}
\caption{Differential rate for the CPC contribution, in arbitrary units, 
including (plain) and excluding (dashed) the $K^{+}K^{-}$ loop. }%
\label{fig:DiffCPC}%
\end{figure}

Replacing the $\cO(p^2)$ amplitude $M_{\pi^{0}PP}^{(2)}(z)$ 
in Eq.~(\ref{eq:fullp2}) with the following parameterization
\begin{equation}
M_{\pi^{0}PP}^{(2)}(z) \to G_{8} m^{2}_K a_{}^{P}(z)~,
\end{equation}
where $G_{8}=9.1 \times 10^{-6}~{\rm GeV}^2$, the CPC differential rate becomes
\begin{equation}
\frac{d\Gamma_{\mathrm{CPC}}}{dz}=\frac{ G_{8}^{2} m_K^5 \alpha^{4} }{ 512\pi^{7} }
\frac{r_{\ell}^{2}}{z}\beta_{\pi}(z)  \beta_{\ell}^{3}(z)  
\left|  a_{}^{\pi}(z)  \mathcal{I}\left(\frac{r_{\ell}^{2}}{z}, 
\frac{r_{\pi}^{2}}{z}\right)  -a_{}^{K}(z)  \mathcal{I}
\left(  \frac{r_{\ell}^{2}}{z},\frac{1}{z}\right)\right|^{2}~.
\end{equation}
We have evaluated this expression using 
several parameterizations of $a_{}^{P}(z)$, 
reported in Table~\ref{tab1}.
The constant one corresponds to the first approximation of Ref.~\cite{Se93}.
The chiral forms, deduced from the $\mathcal{O}\left(p^{2}\right)$ 
amplitudes for $K\rightarrow\pi^{0}\pi^{+}\pi^{-}$ and
$K\rightarrow\pi^{0} K^{+}K^{-}$, are the expressions which yield
to the complete $\cO(p^4)$ calculation of the CPC rate. 
Finally, the Dalitz expression is obtained from 
the experimental fit of the $K_{L}\rightarrow\pi^{0} \pi^{+}\pi^{-}$ 
amplitude~\cite{KMW,CEP}:
\begin{eqnarray}
&& \mathcal{A}\left(  K_{L}\rightarrow\pi^{+}\pi^{-}\pi^{0}\right) = G_{8}%
m_K^{2}a_{}(z)  + \cO\left\{ [p_K\cdot(p_+-p_-)]^2 \right\}~,
\nonumber  \\
&& \qquad a_{}(z)  =0.38+0.13Y_{0}-0.0059Y_{0}^{2}~, \qquad 
Y_{0}=\frac{1}{r_{\pi}^{2}}\left(z-r_{\pi}^{2}-\frac{1}{3}\right)
\end{eqnarray}
neglecting the small $\cO\left\{ [p_K\cdot(p_+-p_-)]^2 \right\}$ terms. 
The resulting branching
ratios for the muon and electron modes are shown in Table~\ref{tab2}.
The differential rate $d\Gamma/dz$ with or without the kaon 
loops and with the Dalitz
parametrization for $a_{}^{\pi}$ is shown in Fig.~\ref{fig:DiffCPC}.
As can be seen,  the impact of the charged kaon loops is quite small.

\medskip

\begin{table}[t]
\begin{center}
\begin{tabular}
[c]{|c|c|c|c|}\hline
& $\text{Constant}$ & $\text{Chiral}$ & $\text{Dalitz}$\\\hline
$a_{}^{\pi}(z)  $ & $\;\;-0.362\;\;$ & $z-r_{\pi}^{2}$ &
$-0.46+2.44z-0.95z^{2}$\\
$a_{}^{K}(z)  $ & $0$ & $\;\;\;z-r_{\pi}^{2}-1\;\;\;$ &
$-$\\\hline
\end{tabular}
\end{center}
\vskip -0.3 true cm
\caption{\label{tab1} 
Phenomenological functions $a_{}^{P}(z)$ 
used in the $K_{L}\rightarrow\pi^{0}P^{+}P^{-}$ vertices.}
\bigskip
\begin{center}
\begin{tabular}
[c]{|cc|c|c|c|c|}\hline
&  & \multicolumn{3}{|c|}{No $KK$} & Chiral $KK$\\\cline{3-6}
&  & $\text{Constant }\pi\pi$ & $\text{Chiral }\pi\pi$ & $\text{Dalitz }\pi
\pi$ & $\text{Dalitz }\pi\pi$\\\hline
$\mu^{+}\mu^{-}$ & \multicolumn{1}{|c|}{$\cB_{\mathrm{CPC}}$} & $3.48\times
10^{-12}$ & $2.26\times10^{-12}$ & $2.62\times10^{-12}$ & $2.81\times10^{-12}%
$\\
& \multicolumn{1}{|c|}{$R_{\gamma\gamma}$} & $4.15\times10^{-6}$ &
$3.80\times10^{-6}$ & $3.58\times10^{-6}$ & $3.66\times10^{-6}$\\
& \multicolumn{1}{|c|}{$R_{abs}$} & $81\%$ & $79\%$ & $77\%$ & $84\%$\\\hline
$e^{+}e^{-}$ & \multicolumn{1}{|c|}{$\cB_{\mathrm{CPC}}$} & $2.28\times10^{-14}$
& $1.48\times10^{-14}$ & $1.75\times10^{-14}$ & $1.80\times10^{-14}$\\
& \multicolumn{1}{|c|}{$R_{\gamma\gamma}$} & $2.73\times10^{-8}$ &
$2.47\times10^{-8}$ & $2.39\times10^{-8}$ & $2.34\times10^{-8}$\\
& \multicolumn{1}{|c|}{$R_{abs}$} & $27\%$ & $28\%$ & $30\%$ & $28\%$\\\hline
\end{tabular}
\end{center}
\vskip -0.3 true cm 
\caption{\label{tab2}
Branching ratio, ratio $R_{\gamma\gamma}$ and absorptive component, 
$R_{abs}=\Gamma_{\mathrm{CPC}}^{abs}/\Gamma_{\mathrm{CPC}}$, 
for the two $K_{L}\rightarrow\pi^{0}\ell^{+}\ell^{-}$ modes, 
as obtained with different phenomenological expressions of
$a_{}^{\pi,K}(z)$.}
\end{table}

Of special interest is the ratio $R_{\gamma\gamma}$, defined in 
Eq.~(\ref{eq:Rgg}), with the two widths computed in terms of the
\textit{same} $a_{}^{P}(z)$ vertex. Concerning the two-photon mode,
this means \cite{CEP}
\begin{equation}
\Gamma\left(  K_{L}\rightarrow\pi^{0}\gamma\gamma\right)  =\frac{G_{8}%
^{2}m_K^{5}\alpha^{2}}{\left(  4\pi\right)  ^{5}}\int_{0}^{\left(  1-r_{\pi
}\right)^{2}}dz\;\beta_{\pi}(z)  \left|  a_{}^{\pi}(z) 
 \mathcal{I}_{\gamma\gamma}\left(  z/r_{\pi}^{2}\right)-a^{K}(z) 
 \mathcal{I}_{\gamma\gamma}(z)  \right|^{2}~,
\end{equation}
with the standard loop function $\mathcal{I}_{\gamma\gamma}$ given in
Ref.~\cite{EPRgg}. 

The phenomenological functions $a_{}^{P}(z)$ act as a
modulation of both the $\ell^{+}\ell^{-}$ and $\gamma\gamma$ spectra, and the
four cases in Table~\ref{tab2}  span a large range of behaviors (this is why 
we kept the constant $a_{}^{P}$, even if this choice is clearly in conflict  
with chiral symmetry and the experimental 
$K_{L}\rightarrow\pi^{0}\gamma\gamma$ spectrum). 
Interestingly, the dynamics of the processes is such that  
$K_{L}\rightarrow\pi^{0} \gamma\gamma$ and
$K_{L}\rightarrow\pi^{0}\ell^{+}\ell^{-}$ 
amplitudes react in a very similar way 
to changes in the distribution of the invariant mass 
of the electromagnetic system 
(i.e. the $\gamma\gamma$ or $\ell^{+}\ell^{-}$ pairs). 
As a result, despite the individual variations of the two widths,
the ratio $R_{\gamma\gamma}$ remains very stable. Note that 
this conclusion is not induced by a complete dominance of the 
$2\gamma$-cut in the $K_{L}\rightarrow\pi^{0}\ell^{+}\ell^{-}$ 
amplitude: $\pi\pi$ and $\pi\pi\gamma$ cuts have sizable
contributions, and the stability of $R_{\gamma\gamma}$ holds 
also for the $K_{L}\rightarrow\pi^{0}e^{+}e^{-}$ amplitude,
which is dominated by the dispersive part.

As is well-known, none of the parameterizations in Table 1 for $a_{}^{P}$ is
able to reproduce both rate and spectrum of the 
$K_{L}\rightarrow\pi^{0}\gamma\gamma$ decay,
and some additional $\mathcal{O}\left(p^{6}\right)$ 
genuine local terms need to be included (see e.g. Refs.~\cite{BDI,Gabbiani}
for a recent discussion).
As shown in Ref.~\cite{BDI}, recent data 
point toward $\mathcal{O}\left(p^{6}\right)$ 
counterterms whose bulk values can be 
interpreted in terms of vector-meson resonances. 
The overall effect of these extra contributions is an 
enhancement of the $K_{L}\rightarrow\pi^{0}\gamma\gamma$
rate, with a spectrum which remains very similar to 
what we have called the Dalitz structure. 
Unfortunately, this information cannot be directly 
translated into a complete $\mathcal{O}\left(p^{6}\right)$ 
analysis of the $K_{L}\rightarrow\pi^{0}\ell^{+}\ell^{-}$ 
amplitude. However, given the stability of $R_{\gamma\gamma}$
in Table~\ref{tab2}, we expect that $R_{\gamma\gamma}$ will 
not change substantially also with the inclusion of these 
extra $\mathcal{O}\left(p^{6}\right)$ terms.
This expectation is further reinforced by 
the observation that the $K_{L}\rightarrow\pi^{0}\gamma\gamma$
spectrum does not show significant evidences of 
two-photon states with $J\not=0$ \cite{BDI}, and by the 
large size of  $R_{\rm abs}$ in Table~\ref{tab2}.
Since we cannot fully determine the structure of the 
$K_{L}\rightarrow\pi^{0}\ell^{+}\ell^{-}$
amplitude at  $\mathcal{O}\left(p^{6}\right)$, 
we shall attribute to our final estimate of  $R_{\gamma\gamma}$
a conservative $30\%$ error, dictated by na\"\i ve chiral power counting.
For the interesting muon mode, this means
\begin{equation}
R_{\gamma\gamma} (\ell=\mu ) = (3.7 \pm 1.7) \times10^{-6}~,
\end{equation}
where the central value has been obtained from the last 
column in Table~\ref{tab2} (Dalitz expression for the pion 
loops and inclusion of the kaon loops).

In order to obtain the predictions of the CPC 
$K_{L}\rightarrow\pi^{0}\ell^{+}\ell^{-}$ widths,
the estimates of $R_{\gamma\gamma}$ need to be combined 
with the experimental results on $\cB\left(  K_{L}\rightarrow\pi^{0}\gamma\gamma\right)$.
The two most recent determinations 
\begin{equation}
\cB\left(  K_{L}\rightarrow\pi^{0}\gamma\gamma\right)^{\exp}=\left\{
\begin{array}
[c]{ll}
\left(  1.36\pm0.03_{\rm stat}\pm0.03_{\rm syst}\pm0.03_{\rm norm}\right)  \times10^{-6} &
\text{\cite{NA48_KLpgg}} \\
\left(  1.68\pm0.07_{\rm stat}\pm0.08_{\rm syst}\right)  \times10^{-6} &
\text{\cite{KTeV_KLpgg}} 
\end{array}
\right.
\end{equation}
leads to the average $\cB\left(  K_{L}\rightarrow\pi^{0}\gamma\gamma\right)^{\exp}
=(1.42\pm0.13)\times10^{-6}$. Using this averaged result, we finally obtain 
\begin{equation}
\begin{array}
[c]{c}%
\cB\left(  K_{L}\rightarrow\pi^{0}e^{+}e^{-}\right)_{\rm CPC}^{0^{++}}=\left(
3.4\pm1.0\right)  \times10^{-14}~, \\
\cB\left(  K_{L}\rightarrow\pi^{0}\mu^{+}\mu^{-}\right)_{\rm CPC}^{0^{++}}=\left(
5.2\pm1.6\right)  \times10^{-12}~. 
\end{array}   \label{eq:CPC00}
\end{equation}

Given the strong helicity suppression, the result in  Eq.~(\ref{eq:CPC00})
is not the dominant CPC contribution for the electron mode. 
As shown in Ref.~\cite{BDI}, the constraints on the $K_{L}\rightarrow\pi^{0}\gamma\gamma$ 
diphoton spectrum (at low invariant mass) leave room for a  $2^{++}$ 
CPC branching ratio, in the electron channel, around or below 
the $10^{-12}$ level. Because of the phase-space and angular-momentum 
suppression,  this translates into a $10^{-14}$ contribution 
from $2^{++}$ in the muon mode, which can thus be safely neglected.

\section{Phenomenological analysis}
The total branching ratios for the two modes can be written as
\begin{equation}
\cB(K_{L}\rightarrow\pi^{0}\ell^{+}\ell^{-})=\left(  C_{\rm mix}^{\ell}
 \pm C_{\rm int}^{\ell}  \left(
\frac{\displaystyle  \mathrm{\operatorname{Im}}\lambda_{t}}{\displaystyle
10^{-4}}\right)  +C_{\rm dir}^{\ell}\left(  \frac{\displaystyle
\mathrm{\operatorname{Im}}\lambda_{t}}{\displaystyle  10^{-4}}\right)^{2}
+C_{\rm CPC}^{\ell}\right)  \times10^{-12},
\end{equation}
with the following sets of coefficients  
(obtained by combining the results of the previous two sections and the 
analysis of Ref.~\cite{BDI} for $K_{L}\rightarrow\pi^{0}e^{+}e^{-}$): 
\begin{equation}
\begin{array}[c]{lcllcl}
 C_{\rm mix}^{e} &=& (15.7\pm0.3) |a_{S}|^{2}~, \qquad & C_{\rm mix}^{\mu} &=& (3.7\pm0.1) |a_{S}|^{2}~, \\
 C_{\rm int}^{e}  &=& (6.2\pm0.3)|a_{S}| ~, & C_{\rm int}^{\mu}  &=& (1.6\pm0.1)|a_{S}|~, \\
 C_{\rm dir}^{e}  &=& 2.4\pm0.2~, &   C_{\rm dir}^{\mu}  &=& 1.0\pm0.1~, \\
C_{\rm CPC}^{e} &\approx& 0~, & C_{\rm CPC}^{\mu}  &=& 5.2\pm1.6~, 
\end{array}
\end{equation}
where  $|a_{S}|= 1.2 \pm 0.2$ and 
$\left(\operatorname{Im}\lambda_{t}/10^{-4}\right)_{\rm SM} = 1.36 \pm 0.12$.
The Standard Model predictions are then
\begin{equation}
\begin{array}{ll}
\cB(K_{L}\rightarrow\pi^{0}e^{+}e^{-})=\left(3.7_{-0.9}^{+1.1}\right)\times10^{-11} &
\big[  \left(1.7_{-0.6}^{+0.7})\times10^{-11}\right ) \big]~, \\
\cB(K_{L}\rightarrow\pi^{0}\mu^{+}\mu^{-})=\left(  1.5\pm0.3\right) \times 10^{-11}   {~}^{~} & 
\big[ ~ (1.0 \pm 0.2) \times 10^{-11})~\big]~, \label{eq:SMrates}
\end{array}
\end{equation}
for positive [negative] interference. As already mentioned, several 
theoretical arguments point towards the constructive interference 
scenario \cite{BDI,GD}. We report the destructive solution only 
for completeness, since we cannot prove yet in a model-independent way 
that this solution is ruled out.

\begin{figure}[t]
\begin{center}%
\[
\includegraphics[width=15.0cm,height=5.2cm]{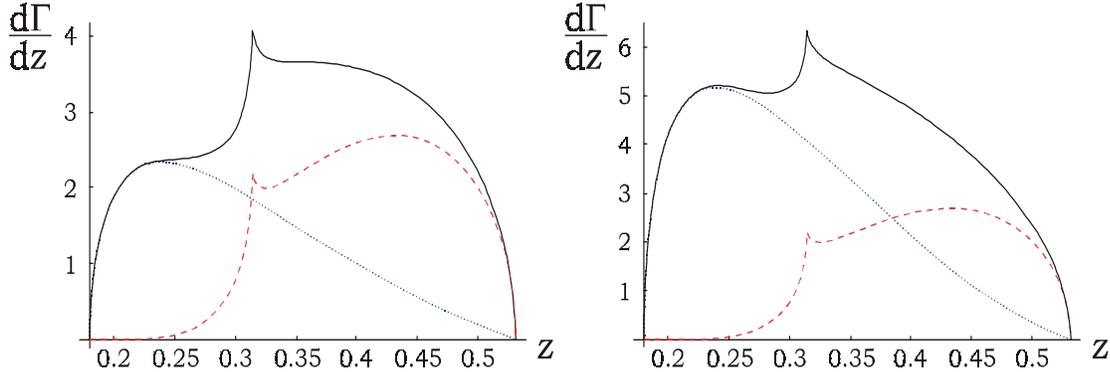}  \hspace{1cm}%
\]
\end{center}
\caption{Differential rate for $K_{L}\rightarrow\pi^{0}\mu^{+}\mu^{-}$, in arbitrary units, 
for destructive (left) and constructive (right) interference. In each plot the three curves 
correspond to CPC (dashed), CPV (dotted) and  total contribution (full).}%
\label{fig:DiffTot}%
\end{figure}

The total differential rate and its CPC and CPV components are shown in
Fig.~\ref{fig:DiffTot}  for the muon mode  (the electron case is less interesting, 
being completely dominated by the CPV part). Although the differential CPC 
component is not predicted as accurately as the total one  
(by means of $R_{\gamma\gamma}$), it clearly emerges that the low $\mu^{+}\mu^{-}$ 
invariant mass region is dominated  by the  CPV component.
This conclusion can be considered as a model-independent statement, 
at least at a qualitative level: 
it follows from the na\"\i ve distribution of $|\mu^+\mu^-\rangle$
final states with $J^{\rm CP}=1^{\pm\pm}$ (CPV) or $0^{++}$ (CPC), and 
it is reinforced by the experimental evidence of the  $K_{L} \rightarrow\pi^{0}\gamma\gamma$ diphoton
spectrum \cite{NA48_KLpgg,KTeV_KLpgg}. For this reason, in future high-statistics 
experiments a cut of the high-$z$ region 
could be used as a powerful tool to reduce and control the CPC contamination of the signal.
At a  quantitative level, employing the Dalitz shape of the spectrum,
we find that cutting the events with $m_{\mu\mu} > 2m_\pi$ let us suppress 
about $90\%$ of the CPC contribution, while the interesting CPV term is reduced  
only by $40\%$ (i.e. the CPC contamination of the total
rate drops to less than $10\%$).  

The different shape of CPC and CPV spectra implies a suppression 
of their interference in the CPV distributions, such as the forward-backward (FB) asymmetry. 
In particular, we find that the normalized FB asymmetry integrated over the full 
spectrum do not exceed the $\cO(10\%)$ level. Thus we do not consider this observable
particularly promising for near-future (low-statistics) experiments.
On the other hand, it is clear that in a long term (high-statistics) perspective
this observable, as well as the other asymmetries discussed in Ref.~\cite{Diwan},
would provide a useful tool to reduce the theoretical error on the 
long-distance components of the $K_{L} \rightarrow\pi^{0}\mu^+\mu^-$ amplitude.

\medskip

The present experimental limits on the branching ratios of the two modes 
can be translated into bounds on $\operatorname{Im}\lambda_{t}$, although 
we are clearly far from the level necessary to perform  precision tests of the SM.
We find
\begin{equation}%
\begin{array}
[c]{ll}%
\cB \left(  K_{L}\rightarrow\pi^{0}e^{+}e^{-}\right)  <2.8\times
10^{-10}\;\text{\cite{KTeVelec}} & \rightarrow\left|  \operatorname{Im}%
\lambda_{t}\right|  <1.3\times10^{-3}~,\\
\cB \left(  K_{L}\rightarrow\pi^{0}\mu^{+}\mu^{-}\right)  <3.8\times
10^{-10}\;\text{\cite{KTeVmuon}} & \rightarrow\left|  \operatorname{Im}%
\lambda_{t}\right|  <2.1\times10^{-3}~,
\end{array} \label{Imlim}
\end{equation}
where the results have been obtained without any assumption on the sign of 
the interference. It is worth to stress that the limit derived from the 
muon mode is quite close to the electron one. This happens because the 
phase-space suppression of the former is partially compensated by the 
enhancement of the helicity-suppressed terms in the (direct-CPV) axial-current 
contribution.

\subsection{New-physics sensitivity}

\begin{figure}[t]
\begin{center}%
\[
\includegraphics[width=12cm]{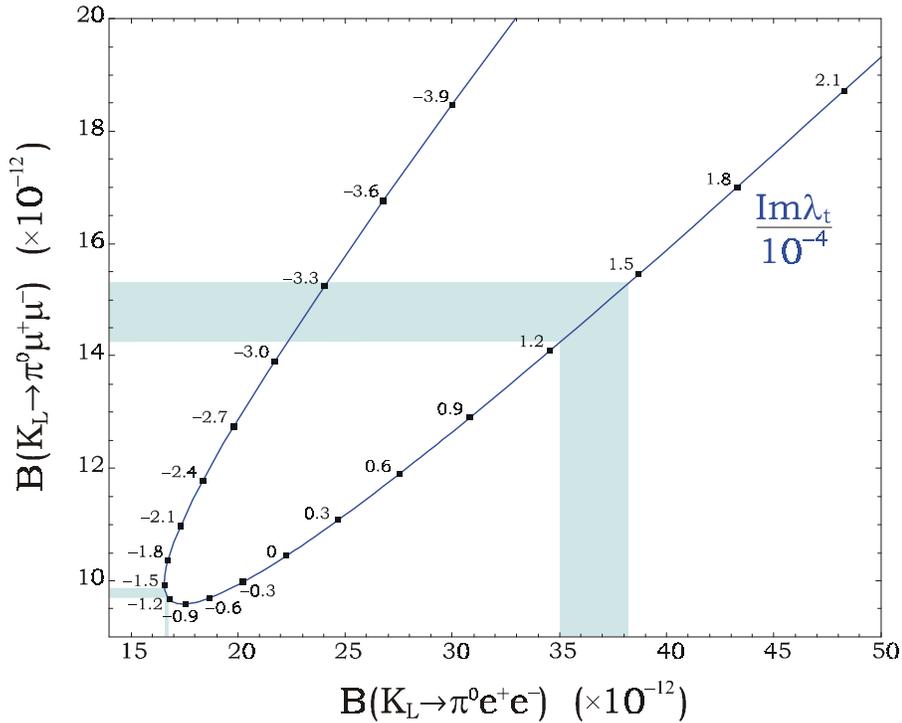}  \hspace{1cm}%
\]
\end{center}
\caption{Central values of the two $\cB(K_{L}\rightarrow\pi^{0}\ell^{+}\ell^-)$
for various values of $\operatorname{Im}\lambda_{t}$. Negative values
stand for destructive interference (see text).}
\label{fig:SMpred}%
\end{figure}

As can be deduced by the comparison of Eq.~(\ref{Imlim}) and  Eq.~(\ref{eq:SMrates}), 
we still have to wait a few years before 
any experiment will be able to probe the two $K_{L}\rightarrow\pi^{0}\ell^{+}\ell^-$
modes at the SM rate. However, we recall that these two transitions
are particularly sensitive to physics beyond the SM and their branching 
ratios could well be modified (and possibly enhanced) with respect to their SM expectations 
(see e.g. Refs.~\cite{BCIRS} and references therein). In such a case, the 
combined information on the two rates would provide a very powerful
tool to identify a deviation from the SM, and also to distinguish 
various new-physics scenarios.

To illustrate the combined discovery potential of the two modes, it is useful 
to draw the curve in the $\cB(K_{L}\rightarrow\pi^{0}\mu^{+}\mu^-)$--
$\cB(K_{L}\rightarrow\pi^{0} e^{+} e^-)$ plane obtained by a variation of  
$\operatorname{Im}\lambda_{t}$, as shown in Fig.~\ref{fig:SMpred}.
The distance between the positive and negative $\operatorname{Im}\lambda_{t}$ 
branches is generated by the different weights of the CPV
contributions for the two modes, which in turn arise from the helicity-suppressed terms 
proportional to $P_{0}(z)$ and $A_{0}(z)$ in Eq.~(\ref{eq:Gamma_CPV}).  
In the limit where we neglect all errors but the one on $\operatorname{Im}\lambda_{t}$, 
as done in Fig.~\ref{fig:SMpred} for illustration, the SM corresponds to one small segment  
(or two considering also the negative interference) of this curve. 
Any other point along the curve corresponds to non-standard scenarios where the 
new-physics effect  can be re-absorbed into a re-definition of $\operatorname{Im}\lambda_{t}$. 
Finally, the region outside the curve corresponds to 
non-standard scenarios with different weights of vector and
axial-vector contributions.

\begin{figure}[t]
\begin{center}%
\[
\includegraphics[width=12cm]{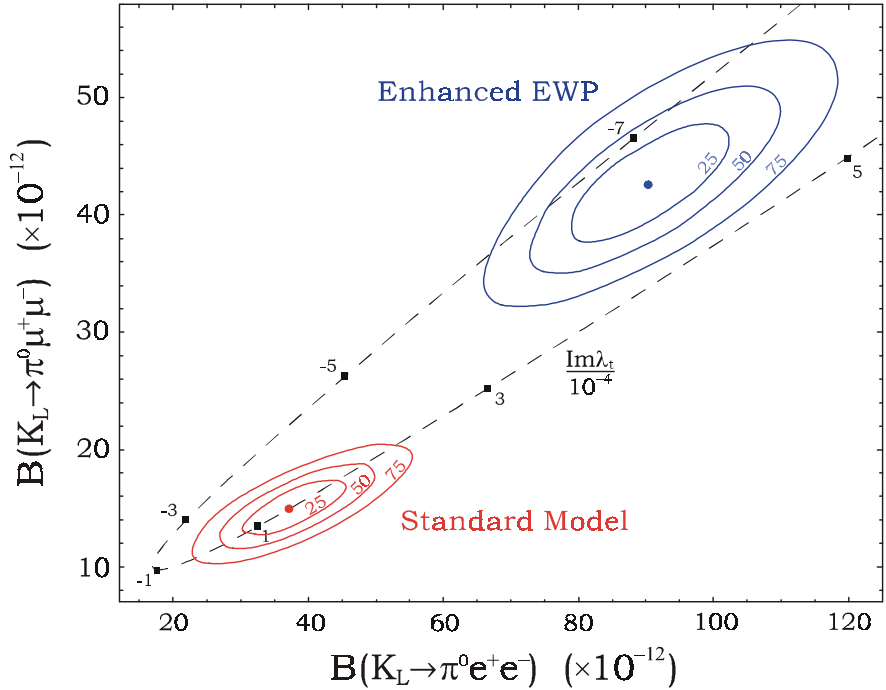}  \hspace{1cm}%
\]
\end{center}
\caption{25\%, 50\% and 75\% confidence-level regions for the Standard Model
(assuming positive interference) and the enhanced-electroweak-penguin model of Ref.~\cite{BFRS},
taking into account all the present uncertainties.}%
\label{fig:NewPhys}%
\end{figure}

To discuss a concrete example, we shall consider in particular the scenario with 
enhanced electroweak penguins recently discussed in Ref.~\cite{BFRS}.
In this framework new physics does not modify the value of  $\operatorname{Im}\lambda_{t}$,
but leads to a modification of the short-distance Wilson coefficients 
$\bar {y}_{7V,A}\equiv y_{7V,A}/\alpha(M_{Z})$. In particular, their values are changed
to $\left(\bar{y}_{7A},\bar{y}_{7V}\right)=\left(  -3.2,0.9\right)$,
whereas the SM case corresponds to $\left(-0.68,0.73\right)$. 
Given the dependence of $C_{\rm int, dir}^{\ell}$ from $\bar{y}_{7V,A}$
\begin{equation}
\begin{array}
[c]{ll}%
C_{\rm int}^{e}=8.91\; \bar{y}_{7V}~|a_S|, & C_{\rm dir}^{e}=2.67\left(  \bar{y}_{7V}%
^{2}+\bar{y}_{7A}^{2}\right)  ~,\\
C_{\rm int}^{\mu}=2.12\; \bar{y}_{7V}~|a_S|, & C_{\rm dir}^{\mu}=0.63\left(  \bar{y}%
_{7V}^{2}+\bar{y}_{7A}^{2}\right)  +0.85\; \bar{y}_{7A}^{2}~,
\end{array}
\end{equation}
we can easily translate this information into the corresponding predictions for the two
branching ratios:
\begin{align*}
\cB(K_{L}  &  \rightarrow\pi^{0}e^{+}e^{-})^{\rm NP}_{\rm EWP } = \left(  9.0\pm1.6\right)
\times10^{-11}~,\\
\cB(K_{L}  &  \rightarrow\pi^{0}\mu^{+}\mu^{-})^{\rm NP}_{\rm EWP} = \left( 4.3\pm0.7\right)
\times10^{-11}~.
\end{align*}
Both decay widths turn out to be dominated by the direct CPV component,
with a small (20\%--30\%) dependence from the interference sign. 
Since the helicity suppressed effects are proportional to the enhanced
$\bar{y}_{7A}$, the relative deviation with respect to the SM prediction is larger
in the muon case. As shown in  Fig.~\ref{fig:NewPhys}, this scenario 
could clearly be distinguished from the SM case even taking into account
all the present parametric uncertainties.  Note that, as a consequence of 
the non-standard axial/vector ratio, the central point of this new-physics
scenario is outside the $\operatorname{Im}\lambda_{t}$ curve.

\section{Conclusions}
We have presented a new comprehensive study of the rare FCNC process 
$K_L \to \pi^0\mu^+\mu^-$. Our main task has been a systematic 
analysis of the two-photon CP-conserving contribution,
which represents the most delicate ingredient necessary 
to obtain a reliable SM  prediction for the total decay rate.
By means of an explicit two-loop calculation within CHPT, we have shown 
that this contribution can be predicted with reasonable accuracy in terms of the 
$K_L\to\pi^0 \gamma\gamma$ width. Taking into account the  experimental 
data on the latter \cite{NA48_KLpgg,KTeV_KLpgg}
and the recent observation of the $K_{S}\rightarrow\pi^{0}\ell^{+}\ell^{-}$ 
transitions \cite{NA48ee,NA48mm}, we have been able to present the first 
complete SM prediction for the total branching ratio:
\begin{equation}
\cB( K_{L} \rightarrow\pi^{0}\mu^{+}\mu^{-} )^{\rm SM} = \left(1.5\pm0.3\right) \times 10^{-11}~.
\end{equation}
This is composed by approximately equal CP-conserving, 
indirect-CP-violating and direct-CP-violating components (including the interference).  
The sizable uncertainty which affects this prediction is a parametric error 
which reflects the present poor experimental knowledge of the $K_S \to\pi^0 \ell^+ \ell^-$ rates. 
The irreducible theoretical error, due to the long-distance 
CP-conserving amplitude, is only at the $10\%$ level and, as we have shown, 
can be further reduced by suitable kinematical cuts. 
  
Beside the good theoretical control of the decay rate,
this channel is particularly interesting because of a different 
short-distance sensitivity compared to the other golden 
rare $K_L$ decays, namely $K_{L} \rightarrow\pi^{0}e^{+}e^{-}$
and $K_{L} \rightarrow\pi^{0}\nu\bar{\nu}$. 
As we have shown, the  helicity-suppressed terms distinguish the relative 
weights of vector- and axial-current contributions in 
$K_{L} \rightarrow\pi^{0}e^{+}e^{-}$ and $K_{L} \rightarrow\pi^{0}\mu^{+}\mu^{-}$.
As a result, the combination of the two decay widths could provide a very powerful
tool to falsify the SM and also to distinguish among different new-physics
models.

In summary, the $K_{L} \rightarrow\pi^{0}\mu^{+}\mu^{-}$
decay represents a very interesting candidate for future precision studies 
of CP violation and new physics in $\Delta S=1$ transitions.
This mode has complementary virtues and disadvantages compared 
to the other rare $K$ decays, and it should be 
seriously considered in view of future high-statistics experiments
in the kaon sector.

\section*{Acknowledgments}
It is a pleasure to thank Augusto Ceccucci, Giancarlo D'Ambrosio, 
Gerhard Ecker, and Cri\-sitina Lazzeroni for useful comments and discussions. 
We are also grateful to David Greynat and Eduardo de Rafael for the
correspondence about Ref.~\cite{GD}.
This work  has been partially supported 
by IHP-RTN, EC contract No.\ HPRN-CT-2002-00311 (EURIDICE).

\end{document}